# THERMAL ANNEALING STUDY OF SWIFT HEAVY-ION IRRADIATED ZIRCONIA

**Jean-Marc COSTANTINI***, Commissariat à l'Énergie Atomique/Saclay, DMN/SRMA, F-91191 Gif-sur-Yvette Cedex, France.

**Andrée KAHN-HARARI,** Laboratoire de Chimie Appliquée de l'Etat Solide, UMR-CNRS 7574, Ecole Nationale Supérieure de Chimie de Paris, 11 Rue Pierre et Marie Curie, F-75231 Paris Cedex 05, France.

**François BEUNEU,** Laboratoire des Solides Irradiés, CEA-CNRS, École Polytechnique, F-91128 Palaiseau Cedex, France.

**François COUVREUR,** Commissariat à l'Énergie Atomique/Saclay, DMN/SEMI, F-91191 Gif-sur-Yvette Cedex, France.

## ABSTRACT

Sintered samples of monoclinic zirconia ($\alpha$-ZrO$_2$) have been irradiated at room temperature with 6.0-GeV Pb ions in the electronic slowing down regime. X-ray diffraction (XRD) and micro-Raman spectroscopy measurements showed unambiguously that a transition to the 'metastable' tetragonal phase ($\beta$-ZrO$_2$) occurred at a fluence of $6.5 \times 10^{12}$ cm$^{-2}$ for a large electronic stopping power value ($\approx 32.5$ MeV $\mu$m$^{-1}$). At a lower fluence of $1.0 \times 10^{12}$ cm$^{-2}$, no such phase transformation was detected. The back-transformation from $\beta$- to $\alpha$-ZrO$_2$ induced by isothermal or isochronal thermal annealing was followed by XRD analysis. The back-transformation started at an onset temperature around 500 K and was completed by 973 K. Plots of the residual tetragonal phase fraction deduced from XRD measurements versus annealing temperature or time are analyzed with first- or second-order kinetic models. An activation energy close to 1 eV for the back-transformation process is derived either from isothermal annealing curves, using the so-called "cross-cut" method, or from the isochronal annealing curve, using a second-order kinetic law. Correlation with the thermal recovery of ion-induced paramagnetic centers monitored by EPR spectroscopy is discussed. Effects of crystallite size evolution and oxygen migration upon annealing are also addressed.

---

* **Corresponding author :** jean-marc.costantini@cea.fr

# I. INTRODUCTION

Zirconium alloys (Zircaloy) are currently used as fuel cladding in pressurized water nuclear reactors. Under operating conditions, an oxide layer is formed at the alloy surface, thereby modifying the mechanical properties of the cladding. It is thus of paramount importance to understand the basic mechanisms of phase transformations in zirconia ($ZrO_2$) under irradiation with heavy ions, either at low energies in the nuclear stopping power regime, corresponding to the neutron-induced recoils, or at high energies in the electronic stopping power regime, corresponding to the fission fragments. Zirconia presents three allotropic forms. In standard temperature and pressure conditions, the thermodynamically stable form of zirconia is the monoclinic (m) one ($\alpha$-$ZrO_2$) with a structure belonging to the $P2_1/c$ space group. With increasing temperature, zirconia first transforms into a tetragonal (t) form ($\beta$-$ZrO_2$) above 1273 K, and then into a cubic (c) form ($\gamma$-$ZrO_2$) at about 2573 K. The $\beta$ and $\gamma$ phases belong respectively to the $P4_2/nmc$ and $Fm3m$ space groups.[1] The $\beta \rightarrow \alpha$ phase transition on cooling is known as a martensitic transformation. However, many experiments are reported on retaining metastable tetragonal zirconia at room temperature.[1]

It was shown in previous studies that the m→t transformation may occur either with low-energy or with high-energy heavy ion irradiations. Such a transformation occurs under 300-keV Kr and 340-keV Xe ion irradiations at 120 K,[2] or 400-keV O, 400-keV Xe and 800-keV Bi ion irradiations at room temperature (RT),[3] in the nuclear slowing down regime. We have shown that this phase transition also occurs after 170-MeV Kr ion irradiation in the electronic slowing down regime, with an electronic stopping power ($S_e$=-$(dE/dx)_e$) ranging between 15.6 and 18.3 MeV $\mu m^{-1}$.[4] However, X-ray diffraction (XRD) analysis showed that the transformation was not complete and limited to 90% t-form fraction even for a high ion fluence of $1.0 \times 10^{14}$ $cm^{-2}$. A double-impact model was needed to account for the transformation kinetics corresponding to a sigmoid plot of the tetragonal phase fraction versus fluence. At a fluence of $1 \times 10^{12}$ cm$^{-2}$, no tetragonal phase was detected. Further on, the threshold of electronic stopping power for such a phase transformation induced by electronic excitations was found to be close to 13 MeV μm$^{-1}$ by using swift heavy ions (Ni, Ge, I) in the 1 to 10 MeV uma$^{-1}$ energy range,[5] corresponding to fission fragment irradiations.

In order to study the thermal evolution of such 'metastable' tetragonal ZrO$_2$ form retained at room-temperature after high-energy heavy ion irradiations, we performed isochronal and isothermal annealing experiments, analyzed by XRD measurements. We show that the t→m back-transformation obeys a thermally-activated second-order kinetics versus annealing time with a corresponding activation energy close to 1 eV.

## II. EXPERIMENTAL PROCEDURE

Pure monoclinic zirconia ceramics (m-ZrO$_2$) were used with a mass density around 95% of the theoretical one (5.83 g cm$^{-3}$). Such high-density pellets were obtained by hot-pressing sintering of high-purity (99.97%) powders during 15 min at 1850°C and 2 MPa in a graphite furnace. Bulky sintered ceramics were machined into half-discs of 12-mm diameter and 1-mm thickness, and annealed during 24 h at 800°C in air, in order to obtain the correct O/Zr stoechiometry, then polished down to 0.5 μm.

Samples were irradiated at GANIL (Caen, France) with 29-MeV uma$^{-1}$ Pb ions in the electronic slowing down regime at only two fluences of $1.0 \times 10^{12}$ and $6.5 \times 10^{12}$ cm$^{-2}$. Irradiations were performed under normal incidence at RT and high vacuum. The electronic energy loss ($S_e$=32.5 MeV μm$^{-1}$) was much larger than the nuclear energy loss ($S_n$=18 keV μm$^{-1}$) at the ion entrance energy, as calculated with the SRIM2000/TRIM96 code.[6] The ion flux was less than $5 \times 10^8$ cm$^{-2}$ s$^{-1}$ in order to avoid electrical charging and macroscopic thermal effects at the most. All as-irradiated pellets turned from white to black.

X-ray powder diffraction spectra were recorded on a D-5000 SIEMENS equipment using Co K$\alpha$ radiation ($\lambda$=1.7890 Å), in the reflection Bragg-Brentano configuration. The analyzed depth, about 23 μm, was clearly smaller than the 159-μm ion projected range. Micro-Raman spectroscopy was used to characterize samples with a T-64000 confocal Jobin-Yvon spectrometer using the 514.5 nm spectrum line of an argon laser beam with a beam spot of 1 μm$^2$. A laser output lower than 300 mW was used to avoid sample damage. The analyzed depth was around 2.5 μm on the basis of an absorption coefficient of 0.4 μm$^{-1}$.[7] EPR measurements were carried out at RT with a computerized Bruker ER 200D X-band spectrometer operating at 9.6 GHz. In the latter analysis, the whole sample volume was probed.

XRD measurements were carried out on a sample irradiated at 6.5x10$^{12}$ cm$^{-2}$ and isochronally annealed in air in an alumina crucible during 30 min, at successive increasing temperatures up to 1223 K ($\pm$ 10 K), below the normal m$\rightarrow$t transformation temperature. Other equivalent samples (irradiated at 6.5x10$^{12}$ cm$^{-2}$) were analyzed after isothermal annealing in the same conditions at 723 K, 773 K, 823 K and 873 K ($\pm$ 10 K) during various successive time intervals. EPR spectra were recorded on the isochronally annealed samples irradiated at 1.0x10$^{12}$ or 6.5x10$^{12}$ cm$^{-2}$.

**III. RESULTS**

XRD powder patterns show a clear transformation of monoclinic zirconia (Fig. 1(a)) induced by 6.0-GeV Pb ion irradiation in a sample irradiated at 6.5x10$^{12}$ cm$^{-2}$ (Fig. 1(c)), similar to our previous results with 170-MeV Kr ion irradiation.[4] Due to the slight c/a distortion ($\approx$1.02),[1] XRD data cannot discriminate between the tetragonal and cubic forms. Unambiguous evidence of a m$\rightarrow$t zirconia phase transition is given by the micro-Raman spectra of a sample irradiated at 6.5x10$^{12}$ cm$^{-2}$, where the characteristic lines at 150 and 268

cm$^{-1}$ of the t-form[4] appear clearly (Fig. 2(c)). This is not seen in the XRD powder patterns and Raman spectra of samples irradiated at 1.0x10$^{12}$ cm$^{-2}$ (Figs. 1(b)-2(b)) which are similar to those of a virgin sample (Figs. 1(a)-2(a)). Characteristic XRD and Raman lines of the residual m-form in the transformed samples, irradiated at 6.5x10$^{12}$ cm$^{-2}$, are broadened with respect to the reference spectrum of virgin sample. Moreover, some line intensities are clearly decreased. However, quantitative treatment of Raman spectra is difficult when looking for the residual tetragonal phase fraction ($X_t$).[8] Therefore θ-2θ XRD was used for quantitative analysis as in our previous study.[4]

XRD patterns were recorded in the selective 30°-45° 2θ range. In agreement with Raman data, the (101) reflection of t-form appears in between the (11-1) and (111) reflections of m-form (Fig. 2). As previously reported,[4] we assessed $X_t$ by using Garvie and Nicholson's formula:[9]

$$X_t = \frac{I(101)_t}{I(11\bar{1})_m + I(101)_t + I(111)_m} \quad (1),$$

where $I$(hkl)$_i$ is the area of the (hkl) peaks of phase i obtained by using the Socabim Profile decomposition software with Pseudo-Voigt profiles. Such a formula was satisfactorily tested for known powders with a variable tetragonal to monoclinic phase volume ratio.[10] For all samples irradiated at 6.5x10$^{12}$ cm$^{-2}$, we found that the m→t transformation is not complete after irradiation: i.e. $X_t^0$ = 64%.

X-band EPR spectra of these samples exhibit complex structures, due to random orientation of grains in polycrystalline pellets (Fig. 3(b)). Similar spectra are found for samples irradiated at 1.0x10$^{12}$ cm$^{-2}$ (Fig. 3(a)), clearly showing that the same paramagnetic centers are produced in all irradiated samples regardless of fluence. Accordingly, all samples turned from white to black after irradiation. XRD patterns of a sample irradiated at $6.5 \times 10^{12}$ cm$^{-2}$ were recorded after isochronous annealing at increasing temperatures and isothermal annealing at increasing time (Fig. 4). Characteristic EPR lines of the paramagnetic centers vanished after annealing at 723 K in all irradiated samples.

## IV. DATA ANALYSIS

The R-ratio of the residual t-form fraction in the isochronously annealed sample ($X_t$), normalized to the initial as-irradiated sample value ($X_t^0$=64%), is then plotted versus annealing temperature (Table I and Fig. 5(a)). The relative error on R is below 10%. The R-ratio is also plotted versus annealing time, for isothermal annealing at various temperatures (Fig. 6). It comes out that R begins to decrease at an onset temperature of 500 K and tends to zero by 973 K. The full width at half maximum (FWHM) of (11-1)$_m$ and (111)$_m$ reflections decreased respectively from about 1.2° and 0.8° to a value close to 0.3° after annealing at 973 K, where the back-transformation was completed (Table I). It further decreased down to about 0.2° for both reflections after annealing at 1223 K. The latter value is close to FWHM of the untransformed samples (irradiated at $1.0 \times 10^{12}$ cm$^{-2}$) and reference virgin samples. The FWHM of (101)$_t$ reflection decreased weakly from 0.8° to 0.6° at 823 K, where the peak intensity became very weak (R≈0.2) (Table I). FWHM values, normalized to the as-irradiated ones, are also plotted versus annealing temperature (Fig. 5(b)). Crystallite sizes of the m-form ($d_m$) and of the t-form ($d_t$) (Table I and Fig. 5(b)) were deduced from the FWHM values according to the Scherrer's formula.[11]

The thermally-induced t→m back-transformation takes place with a single-stage process at a characteristic temperature $T_{1/2}$=700 K, for which 50% of the t-form is transformed into the m-form (Fig. 5(a)). At this temperature, samples become brittle as a result of the 3-5% mass density variation at the t→m transition.[1] Complete restoration of the stable m-form is achieved by 973 K.

**IV.1 Isochronal annealing**

The sigmoid isochronal annealing curve (Fig. 5(a)) was first fitted by using a first-order reaction kinetics law with the following rate equation:

$$dX_t/dt = -\nu X_t \qquad (2).$$

Assuming that:

$$\nu = \nu_0 \exp(-\Delta E/k_B T) \qquad (3),$$

yields a classical thermally-activated behavior for the R-ratio after integration:

$$R = X_t/X_t^0 = \exp[-\nu_0 t_a \exp(-\Delta E/k_B T)] \qquad (4),$$

where $\nu_0$ is a frequency factor, $t_a$=1800 s, the annealing time, $k_B$, the Boltzmann constant, $\Delta E$, the activation energy for recovery, and $T$ the variable annealing temperature. Least-squares fitting according to Eq(4) gives $\Delta E$=0.50 eV.

Since this fit is not fully satisfactory, a second-order reaction kinetics law was also attempted with the following rate equation:

$$dX_t/dt = -\nu' X_t^2 \qquad (5).$$

After integration, it also yields a thermally-activated behavior for the isochronal annealing:

$$R = [1 + \nu_0' X_t^0 t_a \exp(-\Delta E/k_B T)]^{-1} \qquad (6).$$

A better fit is obtained according to Eq(6) with a 0.77 eV activation energy ($\Delta E$).

**IV.2 Isothermal annealing**

A cross-check of such an analysis was then done by fitting the isothermal annealing curves versus time (t) at a given temperature $T_a$ according to the following equations for the first-order reaction kinetics:

$$R = \exp(-t/\tau) \qquad (7),$$

with a decay time:

$$\tau = \nu_0^{-1} \exp(\Delta E / k_B T_a) \qquad (8),$$

and for the second-order reaction kinetics:

$$R = (1 + \alpha t)^{-1} \qquad (9),$$

with:

$$\alpha = \nu_0' X_t^0 \exp(-\Delta E/k_B T_a) \qquad (10).$$

Once again the second-order kinetics law fits the data points much better than the first-order one (Fig. 6).

Moreover, we have used the so-called "cross-cut" method,[12] by plotting the time values corresponding to R=0.25 and R=0.35 on the 4 isotherms (Table II), as a function of the inverse annealing temperature ($1/T_a$) (Fig. 7). It can be shown that the slope of these Arrhenius plots is the activation energy ($\Delta E$).[12] The sets of values corresponding to these two "cross-cuts" yield about the same slopes: i.e. respectively $\Delta E$=1.13 eV and $\Delta E$=1.11 eV.

On the same plot (Fig. 7) are also displayed the values of the time constant ($\alpha^{-1}$) (Table II) in the expression of the second-order kinetics law which is also thermally-activated according to Eq(9). Here again, it yields around the same slope with $\Delta E$=1.11 eV. The exponential decay time ($\tau$) (Table II) corresponding to a first-order kinetics law according to Eq(7) is also shown on this plot (Fig. 7). In the latter case, the discrepancies between the $\tau$-values and the "cross-cut" time values at R=0.35 increase when the annealing temperature increases from 723 K to 873 K (Fig. 7). It yields a somewhat smaller slope with $\Delta E$=0.90 eV.

## V. DISCUSSION

In the case of tetragonal zirconia produced by nuclear collision process with heavy ions in the 100-keV energy range, some authors have claimed that the t→m back-transformation is correlated with bleaching of point defects considered to be oxygen vacancies (F-centers).[13] The t→m back-transformation and recovery of these defects by isochronal annealing (30 min) indeed occurs in a single stage, at about the same characteristic temperature $T_{1/2}$=550 K, with almost total phase transformation and defect bleaching below 700 K.[13] This process was modeled on the basis of second-order kinetics of Frenkel pair recombination.[13]

The latter results are consistent with those observed in c-YSZ. In that case, isochronal annealing curves (30 min) of the paramagnetic $F^+$-like centers produced by 200-MeV Au ion irradiations,[16] at a fluence of $1.0 \times 10^{12}$ cm$^{-2}$, have a characteristic temperature ($T_{1/2}$) at which 50% of defects are bleached out close to 450 K.[14] It is also consistent with other data obtained with 100-MeV C ion irradiations ($S_e$=0.5 MeV μm$^{-1}$) of c-YSZ at low fluences, where $T_{1/2}$ is in the 450-500 K range, showing that this single-stage annealing process, with activation energies ranging between 0.5 and 1 eV, is independent of the ion stopping power.[14,15] At larger fluences, i.e. larger concentrations of color centers, $T_{1/2}$ is shifted to higher values, and a second recovery stage of $F^+$-like centers occurs above 600 K.[15] In this second stage, color centers are transformed into other defects, with full bleaching eventually occurring at 973 K.[15]

However, this second recovery stage is not found in the EPR spectra of the samples studied here, recorded after isochronous annealing, where full color-center bleaching is observed at 723 K in all irradiated samples. A quantitative analysis of these spectra in order to extract the color center concentration is not feasible due to the random orientations of grains. These results entail that the processes involved in the t→m back-transformation cannot be directly correlated with some point defect recovery or reactions, contrary to the case of the m-form induced by nuclear collisions, where color centers are bleached in a single-stage process below 700 K.[13] Indeed, our data show that color centers are fully bleached at 723 K, whereas the complete t→m back-transformation of $ZrO_2$ takes place at 973 K.

Simultaneous analysis of the XRD patterns yields the evolution versus annealing temperature of the line widths of $(101)_t$, $(11\text{-}1)_m$ and $(111)_m$ reflections (Table I). The relative error on FWHM is lower than 5%. In the transformed samples (at $6.5 \times 10^{12}$ cm$^{-2}$), the FWHM of $(11\text{-}1)_m$ and $(111)_m$ reflections exhibit a sigmoid variation similar to that of the R-ratio versus annealing temperature, with a characteristic temperature $T_{1/2} \approx 723$ K (Fig. 5(b)), and a decrease by about a factor 4 up to 973 K (Table I). The FWHM of X-ray lines corresponding to a virgin sample was eventually reached at 1223 K. In contrast, the FWHM of $(101)_t$ reflection showed a much weaker decrease by about 20% up to 823 K (Table I). Point defect contribution to XRD peak broadening can be discarded, since no broadening of $(11\text{-}1)_m$ and $(111)_m$ reflections is observed in the untransformed samples (irradiated at $1.0 \times 10^{12}$ cm$^{-2}$) with respect to virgin samples (Fig. 1(b)). This is in agreement with the Raman lines which are not broadened in these untransformed samples (Fig. 2(b)).

Therefore, we assume that the major contribution to XRD peak broadening in transformed samples should be due to grain coarsening when annealing temperature is increased. Micro-Raman spectrometry data clearly show that the m→t transformation was produced by irradiation on a spot size smaller than 1 μm$^2$. According to Scherrer's formula,[11] we find that the m-form crystallite size ($d_m$) increases from 8-10 nm to 45-55 nm during the annealing process, while the t-form crystallite size ($d_t$) increases weakly from about 13 nm to 15 nm before fading out (Table I, Fig. 5(b)). In the transformed samples, Raman lines of both phases are clearly broadened (Fig. 2(c)), in agreement with literature data on annealed m-$ZrO_2$ powders, with crystallite sizes varying from 5.0 to 120 nm, where broadened Raman spectra are found below a critical value of 15 nm.[17]

This discussion on the annealing processes can be summarized as follows. In the case of heavy ion irradiations in the 100-keV range where damage is mainly due to nuclear collisions, F-centers (oxygen Frenkel pairs) are produced in zirconia, and the t→m back-transformation is related to the recovery of these point defects below 723 K. By contrast, in the case of high-energy ion irradiations, not only Frenkel pairs are produced, but also a small-grained microstructure is induced, and the t→m back-transformation occurring to a much higher temperature (973 K) is attributed to the grain (back)-coarsening during annealing. The formation of an oxide material with nanozied grains (10 nm) induced by electronic excitations was already clearly evidenced by high-resolution transmission electron microscopy (TEM) in the case of yttrium iron garnet ($Y_3Fe_5O_{12}$) irradiated by swift heavy ions.[18] We have interpreted the formation of such a nanostructured material as the recrystallization of the amorphous tracks on the basis of a double impact kinetics and thermal spike model.[19-20]

It is interesting to compare such a t→m transformation to the one occurring upon annealing in oxidizing atmosphere of pure nanocrystalline t-$ZrO_2$ powders produced by a spray pyrolysis technique (at thermodynamic equilibrium).[21] The m-form appeared after annealing during 15 min at 1173 K above a critical $d_t$ value of 22 nm,[21] which is not reached here (Table I). We find instead that $d_m$ increases sharply above the characteristic temperature $T_{1/2}$ =700 K (Fig. 5(b)). The driving force of this solid-state transformation which involves approximately no chemical composition change seems then to arise from other factors like minimization of the interfacial free energy.[22]

This t→m back-transformation involves a thermally-activated process spreading from 500 K to about 1000 K. It is then clearly different from the ideal martensitic transformation between these zirconia phases which is known to be diffusionless and athermal.[1] Moreover, such a process is not directly linked to point defect bleaching occurring below 723 K. The activation energy ($\Delta E$) about 1 eV can be considered as the potential energy barrier height ($\Delta G^*$) to overcome in order to restore the m-form, stable at RT, when starting from the 'metastable' t-form with a higher free energy ($G^t$) than the monoclinic one ($G^m$). It was shown that a variety of 'metastable' t-forms can be obtained either due to the free-energy difference ($G^t$-$G^m$ >0), or to the energy barrier ($\Delta G^*$>0), which leads to contradictions and discrepancies between phase-diagram data.[23] In the latter cases of 'metastable' states, it is linked to sluggish t→m transformation kinetics depending on $\Delta G^*$, which in turn depends on the ceramic microstructure (grain-size distributions, point and extended defects,…), thermal history and stress state. Differential thermal analysis (DTA),[24] Raman spectroscopy and XRD data[21] have indeed shown that the t→m transformation upon heating is more sluggish in the case of smaller crystallite sizes and larger m-form fraction in nanostructured t-$ZrO_2$. The transformation upon cooling is also shifted to a higher temperature in the case of smaller crystallites.[24]

We thus think that this is a nucleation-growth process involving solid state diffusion to the interfaces between the two phases. It can be viewed as a second-order reaction between the two phases, involving the product of both phase concentrations in the rate equation,[25] where the frequency factor ($v'$) in Eq(5) should include thermally-activated diffusion constants. One likely mechanism could be related to oxygen migration, since it was suggested that oxygen vacancies play a key role in the stabilization of $ZrO_2$ allotropes:[26] the t-form is destabilized via annealing of anion vacancies under oxidizing atmosphere.[27] Most likely, oxygen diffusion should also play a major role in the color center bleaching process, just like in the case of ion-irradiated c-YSZ.[14-15]

Complementary investigations such as DTA, TEM, and conductivity measurements are needed for further understanding of the present tetragonal to monoclinic back-transformation of irradiated $ZrO_2$.

## VI. CONCLUSIONS

Using X-ray diffraction combined with micro-Raman spectroscopy, we have shown that a transition from the monoclinic ($\alpha$) to the tetragonal ($\beta$) form of $ZrO_2$ is induced by irradiation with 6.0-GeV Pb ions in the electronic slowing down regime, at a fluence of $6.5 \times 10^{12}$ cm$^{-2}$. No transformation at all was detected at $1.0 \times 10^{12}$ cm$^{-2}$. Thermal restoration of the thermodynamically stable $\alpha$-phase occurred above a 500-K onset temperature and was completed by 973 K. Such a $\beta \rightarrow \alpha$ back-transformation is ruled by second-order reaction kinetics with an activation energy close to 1 eV, during which grain coarsening of the monoclinic phase occurs above 700 K. This transformation is not directly correlated with the thermal bleaching of color centers, which occurs below 723 K, in contrast to previously reported data on $\alpha$-$ZrO_2$ irradiated with lower energy ion irradiations in the nuclear collision regime. However, the back-transformation and point defect recovery might be both controlled by oxygen vacancy diffusion and annealing.

**ACKNOWLEDGMENTS:**

Authors are indebted to Dr. D. Gosset (CEN/SACLAY/DMN/SRMA) for providing the sintered ceramics samples. Dr A. Benyagoub (CIRIL, Caen) is also thanked for his assistance during the irradiation at GANIL.

**Table I:** FWHM of the $(11\text{-}1)_m$ and $(101)_t$ reflections of the m- and t-forms respectively, and crystallite sizes of the m-form ($d_m$) and t-form ($d_t$) at the various isochronous (30 min) annealing temperature ($T_a$).

| $T_a$ (K) | R | FWHM $(11\text{-}1)_m$ (°) | $d_m$ (nm) | FWHM $(101)_t$ (°) | $d_t$ (nm) |
|---|---|---|---|---|---|
| 298 | 1.0 | 1.20 | 8.0 | 0.77 | 12.6 |
| 468 | 1.0 | 1.13 | 8.5 | 0.72 | 13.4 |
| 523 | 0.98 | 1.10 | 8.7 | 0.72 | 13.4 |
| 578 | 0.94 | 1.05 | 9.2 | 0.70 | 13.8 |
| 623 | 0.89 | 0.99 | 9.7 | 0.70 | 13.8 |
| 673 | 0.73 | 0.85 | 11.3 | 0.67 | 14.4 |
| 723 | 0.44 | 0.77 | 12.5 | 0.67 | 14.4 |
| 773 | 0.29 | 0.64 | 15.0 | 0.63 | 15.4 |
| 823 | 0.22 | 0.47 | 20.5 | 0.63 | 15.4 |
| 873 | 0.11 | 0.41 | 23.5 | | |
| 923 | 0.07 | 0.34 | 28.3 | | |
| 973 | 0.0 | 0.28 | 34.4 | | |
| 1223 | 0.0 | 0.21 | 45.8 | | |

**Table II:** Annealing times corresponding to the "cross-cuts" of the isotherms at R=0.25 and R=0.35, first-order ($\tau$), and second-order kinetics ($\alpha^{-1}$) time constants at the various annealing temperature ($T_a$).

| $T_a$ (K) | Time at R=0.25 (h) | Time at R=0.35 (h) | $\tau$ (h) | $\alpha^{-1}$ (h) |
|---|---|---|---|---|
| 723 | 5.0 | 2.77 | 2.9 | 1.74 |
| 773 | 1.40 | 1.0 | 1.0 | 0.55 |
| 823 | 0.49 | 0.30 | | 0.16 |
| 873 | 0.22 | 0.14 | | 0.07 |

**FIGURE CAPTIONS**

**Fig. 1:** XRD θ/2θ powder patterns of sintered monoclinic $ZrO_2$ ceramics: virgin (a, dotted line), as-irradiated with 6.0-GeV Pb ions at $1 \times 10^{12}$ cm$^{-2}$ (b, solid line), and $6.5 \times 10^{12}$ cm$^{-2}$ (c, dashed line).

**Fig. 2:** Micro-Raman spectra of sintered monoclinic $ZrO_2$ ceramics: virgin (a), as-irradiated with 6.0-GeV Pb ions at $1 \times 10^{12}$ cm$^{-2}$ (b), and $6.5 \times 10^{12}$ cm$^{-2}$ (c).

**Fig. 3:** Room-temperature X-band EPR spectra of sintered monoclinic $ZrO_2$ ceramics: as-irradiated with 6.0-GeV Pb ions at $1.0 \times 10^{12}$ cm$^{-2}$ (a) and $6.5 \times 10^{12}$ cm$^{-2}$ (b).

**Fig. 4:** XRD θ/2θ powder patterns of sintered monoclinic $ZrO_2$ ceramics obtained with Co Kα: as-irradiated with 6.0-GeV Pb ions at $6.5 \times 10^{12}$ cm$^{-2}$ (a, solid line), and annealed during 30 min at 623 K (b, dashed line), 773 K (c, dash-dotted line), 923 K (d, dotted line), and 1223 K (e, solid line).

**Fig. 5:** (a) Fraction of the residual tetragonal $ZrO_2$ (open circles) versus annealing temperature. Dashed and solid lines are least-squares fitted curves to the first-order (Eq(4)) and second-order kinetics (Eq(6)), respectively. (b) FWHM of the $(11-1)_m$ reflection (full squares, solid line), normalized to the as-irradiated sample values (left scale), crystallite sizes of the monoclinic phase (right scale) deduced from the $(11-1)_m$ (open squares, dashed line) or $(111)_m$ reflections (open triangles, dashed line), and crystallite sizes of the tetragonal phase deduced from the $(101)_t$ reflection (open diamonds, dashed line), versus annealing temperature. Lines are guides to the eyes.

**Fig. 6:** Isothermal annealing curves of the residual tetragonal $ZrO_2$: 723 K (open circles), 773 K (open up-triangles), 823 K (open down-triangles), and 873 K (open diamonds). Dashed and solid lines are least-squares fitted curves to the first-order (Eq(7)) and second-order (Eq(9)) kinetics, respectively.

**Fig. 7:** Annealing time versus reciprocal annealing temperature: "cross-cuts" of isotherms at R=0.25 (open squares), and at R=0.35 (open triangles), exponential decay time with the first-order ($\tau$) (Eq(7)) (open circles), and second-order kinetics ($\alpha^{-1}$) (Eq(9)) (full circles) decay times. Lines are linear least-squares regressions.

**Fig. 1:**

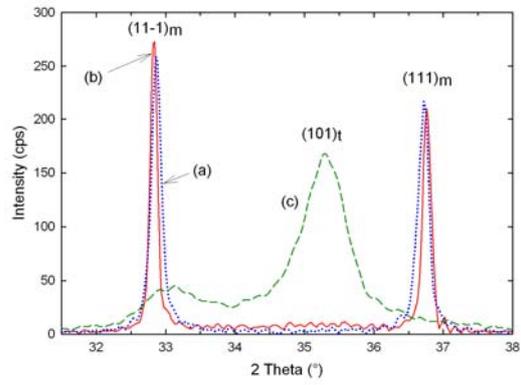

**Fig. 2:**

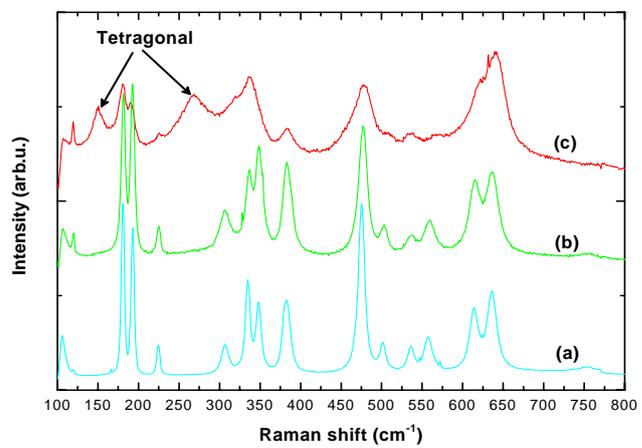

**Fig. 3:**

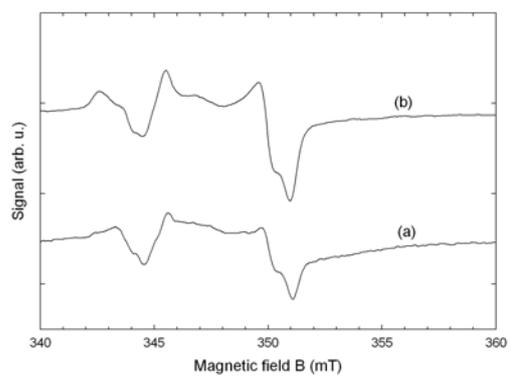

**Fig. 4:**

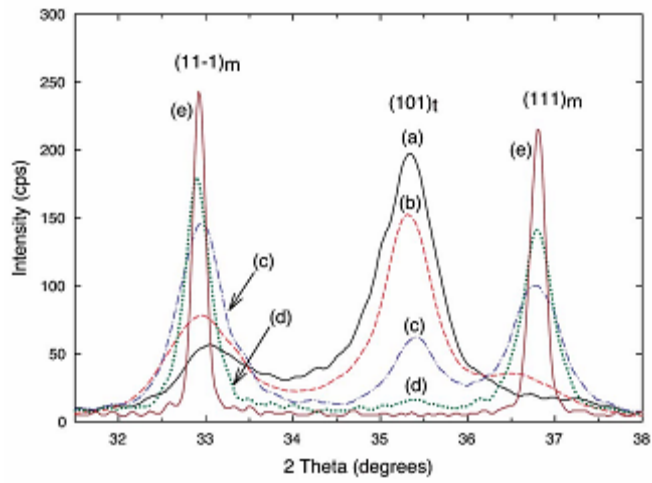

**Fig. 5:**

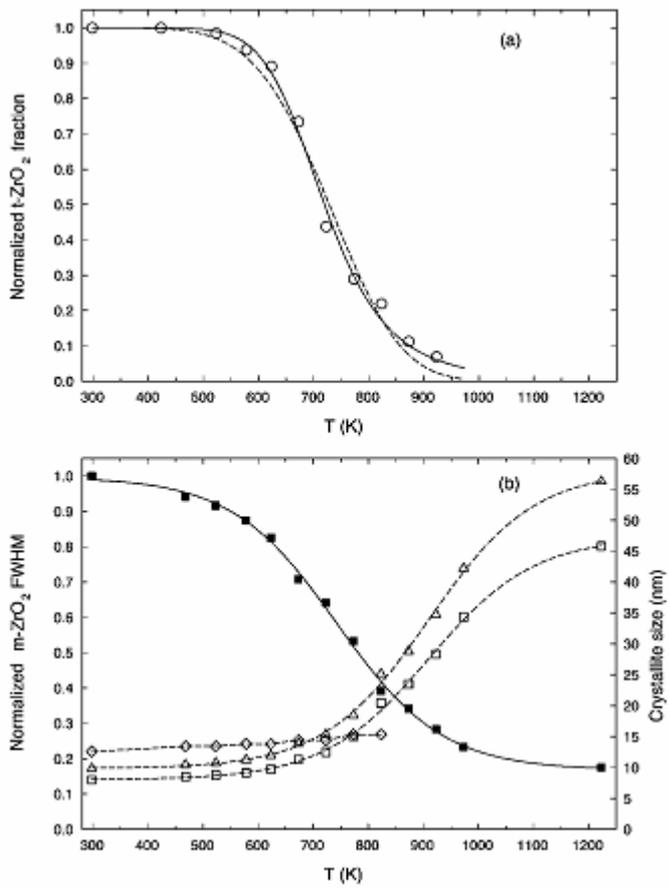

**Fig. 6:**

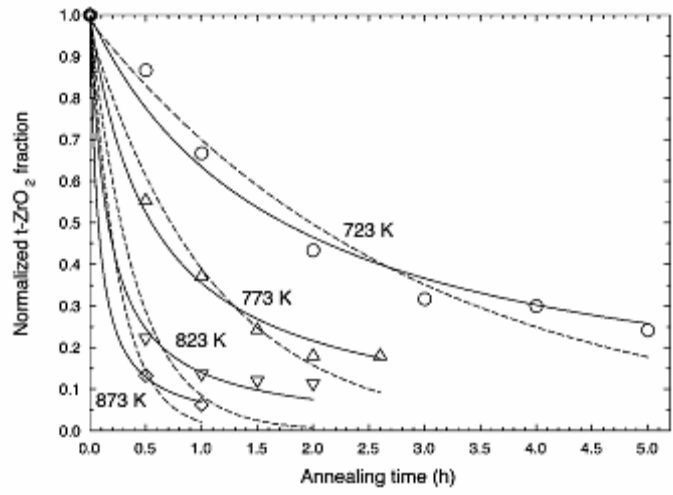

**Fig. 7:**

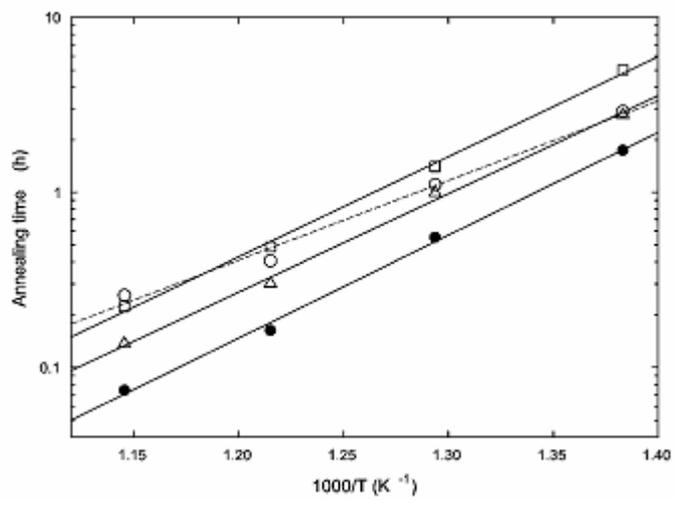